# Comparison of SCAN+$U$ and r$^2$SCAN+$U$ for Charge Density Wave Instability and Lattice Dynamics in CuTe

Seungha Ju[1], Sooran Kim[1,2*]

[1]Department of Physics Education, Kyungpook National University, Daegu 41566, South Korea

[2]KNU G-LAMP Project Group, KNU Institute of Basic Sciences, Kyungpook National University, Daegu, 41566, Korea.

*Corresponding authors: sooran@knu.ac.kr

**Abstract**

Identifying an appropriate exchange-correlation functional and computational conditions is essential for explaining the fundamental physics of materials and predicting their properties. Here, we investigate the performance of the meta-GGA functionals SCAN and r$^2$SCAN, with and without a Hubbard $U$, for describing the charge density wave (CDW) in the quasi-one-dimensional material CuTe. By examining the Te-Te bond modulation, phonon dispersions, and electronic structures, we identify clear differences in how the two functionals capture the structural and dynamical properties of the CDW formation. r$^2$SCAN+$U$ reproduces the experimentally observed Te-chain distortions in the CDW phase and the phonon soft mode at $q_{CDW}$=(0.4, 0.0, 0.5) in the non-CDW phase, whereas SCAN exhibits unphysical phonon behavior. The atomic displacements of the soft mode agree well with the experimental Te modulation. Despite their similar electronic structures and optimized lattice constants, our results demonstrate that r$^2$SCAN is a more suitable choice than SCAN for describing CDW formation and lattice dynamics in CuTe.

# 1. Introduction

Low-dimensional quantum materials have attracted much attention as fertile platforms for discovering and controlling correlated electronic phases such as charge-density waves (CDWs), novel magnetism, and unconventional superconductivity [1–6]. In particular, van der Waals layered compounds and quasi-one-dimensional (quasi-1D) systems provide reduced screening, enhanced nesting, and strong electron-phonon coupling, all of which favor symmetry-lowering lattice instabilities [2,7,8].

The binary chalcogenide CuTe is a prototypical correlated quasi-1D CDW material with a transition temperature $T_{CDW}$=335 K. In its high-temperature non-CDW phase, bulk CuTe crystallizes in an orthorhombic structure (space group *Pmmn*) in which Te atoms form quasi-1D chains along the *a*-axis, while Cu atoms form a buckled square network as shown in Fig. 1(a,b) [9]. Upon cooling below $T_{CDW}$, CuTe undergoes a CDW transition characterized by a 5×1×2 modulation and a CDW vector $q_{CDW}$=(0.4, 0.0, 0.5), as revealed by x-ray diffraction and high-resolution transmission electron microscopy [10]. Angle-resolved photoemission spectroscopy (ARPES) further has shown a momentum-dependent gap opening of order 0.1-0.2 eV on the quasi-1D Te $p_x$ bands [11].

To investigate the properties of CuTe and the origin of its CDW, several first-principles calculations have been reported [10–17]. Zhang *et al*. proposed that the Fermi-surface nesting and electron-phonon coupling contribute to the CDW formation [11]. The role of Coulomb correlation of Cu 3*d* electrons in driving the CDW instability and Te modulations has also been discussed, supported by the appearance of imaginary phonon modes in the DFT+*U* calculations for both bulk and monolayer CuTe [12,13]. Quantum-anharmonic calculations and DFT+*U*+*V* have further been employed to analyze the CDW formation and the associated distortions [14]. This complex origin of the CDW makes CuTe an appealing example for studying CDW mechanisms in correlated low-dimensional systems.

On the other hand, to better capture electronic correlation, the strongly constrained and appropriately normed (SCAN) functional has been introduced as a non-empirical meta-GGA [18,19]. Compared with GGA+$U$ approaches, SCAN can describe important properties of oxides without an adjustable Hubbard $U$, thereby improving the predictive power [20–24]. However, SCAN still suffers from a self-interaction error [25], so an additional on-site $U$ is frequently combined with SCAN [26–29]. Furthermore, SCAN is sensitive to the numerical integration grid, which reduces its utility in large-scale or high-precision calculations [30]. To address this issue, r²SCAN was proposed as a "regularized and restored" version of SCAN that improves numerical stability and grid convergence while retaining the accuracy of SCAN [30]. This makes r²SCAN attractive for systematic studies of complex materials.

A few previous calculations employed SCAN for bulk and monolayer CuTe [31,32]. However, to the best of our knowledge, the 5×1×2 CDW phase of CuTe has not been investigated within SCAN or r²SCAN. Therefore, it is worth investigating whether SCAN and r²SCAN can (i) reproduce the correlation-assisted phonon soft modes in CuTe without explicit $U$, and (ii) reliably describe the Te-chain modulation of the 5×1×2 CDW state.

In this study, we assess the suitability of SCAN and r²SCAN, with and without $U$, for CuTe by analyzing the stability of the CDW modulation, the phonon dispersions with their soft mode eigenvectors, electronic structures, and optimized lattice constants. Through this comparative analysis, we clarify under which conditions r²SCAN+$U$ can properly capture the experimentally observed CDW distortion and phonon instability. We also show that SCAN and r²SCAN, despite yielding similar electronic structures and lattice constants, differ significantly in their description of CDW formation and lattice dynamics.

## 2. Computational Method

All DFT calculations were carried out using the Vienna *ab initio* simulation package (VASP) [33,34]. We employed three exchange-correlation functionals: SCAN [18,19], r²SCAN [30], and PBE (Perdew-Burke-Ernzerhof) [35]. On-site correlation effects in Cu 3$d$ states were included using the Dudarev approach with an effective on-site Coulomb interaction, $U_{\text{eff}} = U - J$ [36]. $U_{\text{eff}}$ values in the range from 1 eV to 13 eV were tested to determine an appropriate value for the system. van der Waals interactions (vdW) were examined using the revised Vydrov-Van Voorhis (rVV10) nonlocal correlation functional [37,38]. The energy cutoff for the plane-waves was set to 400 eV, and atomic positions were relaxed from the initial experimental structure [10] until the residual forces were less than 0.001 eV/Å. We used 20×16×8 and 4×16×4 $k$-point samplings for the non-CDW structure and the 5×1×2 modulated CDW structure, respectively.

Phonon calculations were performed using PHONOPY within the finite displacement (frozen-phonon) method based on the Hellmann-Feynman theorem [39,40]. The dynamical matrices were obtained using a 10×1×2 supercell and a 3×16×4 $k$-point sampling.

## 3. Results and discussion

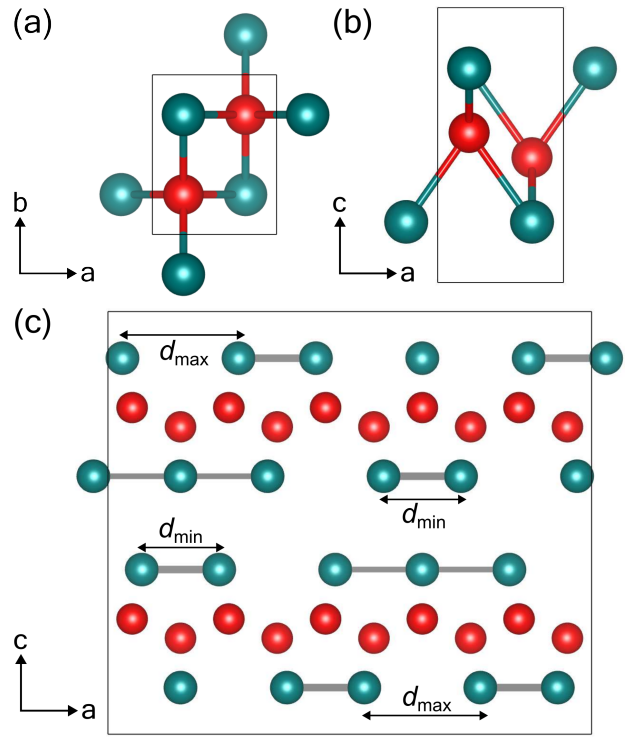

Fig. 1. Crystal structures of the non-CDW and CDW phases of CuTe. Red and emerald spheres represent Cu and Te atoms, respectively. (a) Top view and (b) side view of the non-CDW phase, and (c) the 5×1×2 modulated CDW phase of CuTe. Gray lines show the Te-Te bond modulations; thicker and thinner lines denote Te dimers and trimers, respectively. The longest and shortest Te-Te bonds are labeled $d_{max}$ and $d_{min}$, respectively.

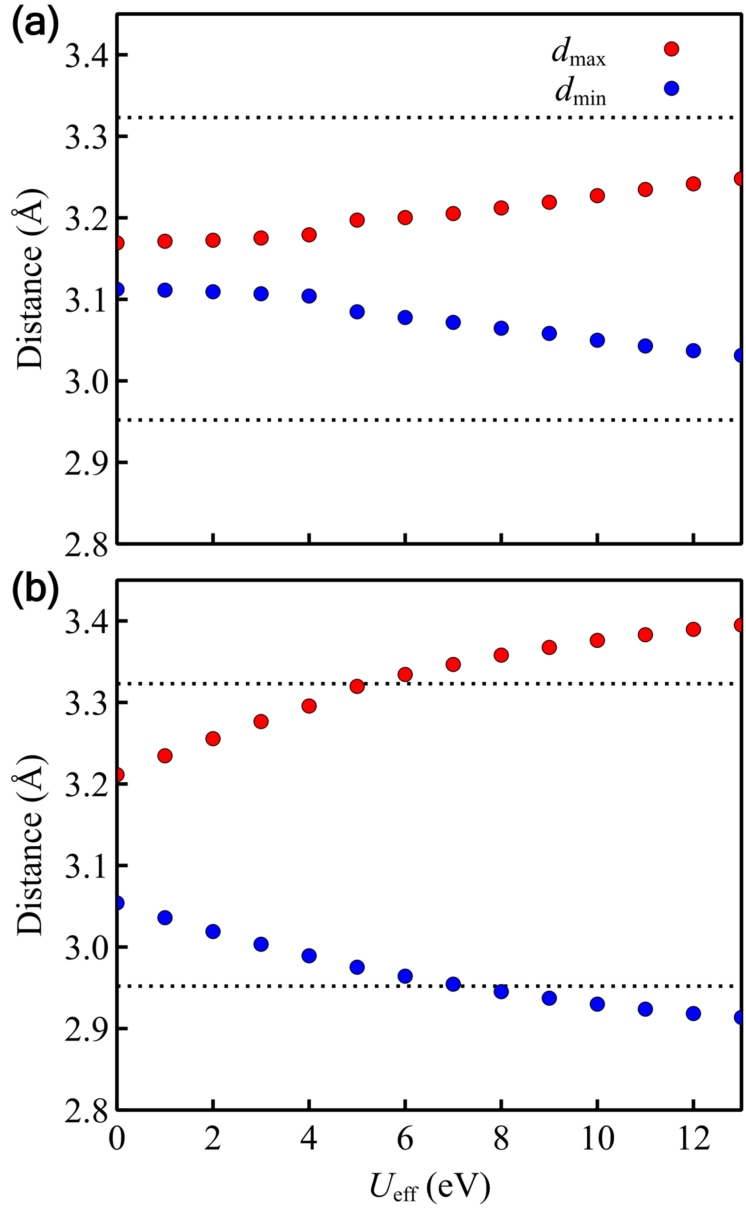

Fig. 2. Calculated Te-Te bond distances as a function of $U_{eff}$ using (a) SCAN and (b) r$^2$SCAN. The horizontal dotted lines at the top and bottom denote experimental $d_{max}$ and $d_{min}$ values, respectively.

To investigate the CDW modulation in CuTe, we calculated the shortest and longest Te-Te distances, $d_{min}$ and $d_{max}$, in the 5×1×2 modulated structure. Figure 2 shows the obtained values using SCAN(+$U$) and r$^2$SCAN(+$U$). Compared to previous PBE calculations [12], even without $U_{\text{eff}}$, both SCAN and r$^2$SCAN already yield the separation between $d_{\text{max}}$ and $d_{\text{min}}$, indicating a finite CDW modulation. This suggests that SCAN and r$^2$SCAN partially capture the local correlation effects that are otherwise modeled by an explicit $U_{\text{eff}}$. The modulation amplitude, $\Delta d = d_{max} - d_{min}$ is significantly larger in r$^2$SCAN than in SCAN. However, for both functionals at $U_{\text{eff}}$= 0 eV, $d_{\text{min}}$ is overestimated while $d_{\text{max}}$ is underestimated compared with experiment [10].

When $U_{\text{eff}}$ is introduced, $\Delta d$ increases for both SCAN and r$^2$SCAN. With r$^2$SCAN, both $d_{\text{max}}$ and $d_{\text{min}}$ almost reach experimental values at $U_{\text{eff}}$ of 5eV. In contrast, the previous PBE+$U$ study reported $\Delta d$ less than 0.1 Å at $U_{\text{eff}}$=5eV, implying a much weaker CDW modulation than observed experimentally. With r$^2$SCAN, a finite $U_{\text{eff}}$ is still required, but the experimental Te modulation is reproduced. We therefore adopt $U_{\text{eff}}$ of 5 eV, for which the difference between calculated and experimental $d_{\text{max}}$ is nearly zero.

Unlike r$^2$SCAN, $\Delta d$ with SCAN increases slowly, and SCAN does not reproduce the experimental modulation within the considered $U_{\text{eff}}$ range. Even at $U_{\text{eff}}$=13 eV, $d_{\text{min}}$ and $d_{\text{max}}$ do not reach the experimental values, and the deviations are even larger than in the PBE+$U$ calculations. It shows that r$^2$SCAN provides a more accurate description of the Te-Te bond modulation and the associated CDW stability than SCAN.

Furthermore, we examined the effects of the rVV10 vdW correction and spin-orbit coupling (SOC). The differences in $d_{\text{max}}$ and $d_{\text{min}}$ between calculations with and without vdW are ~0.015 Å, and SOC changes $d_{\text{max}}$ and $d_{\text{min}}$ by ~0.027 Å. However, they do not alter the overall trends or the relative performance of SCAN and r$^2$SCAN. These insignificant effects of vdW interactions and SOC are consistent with previous studies [12,13].

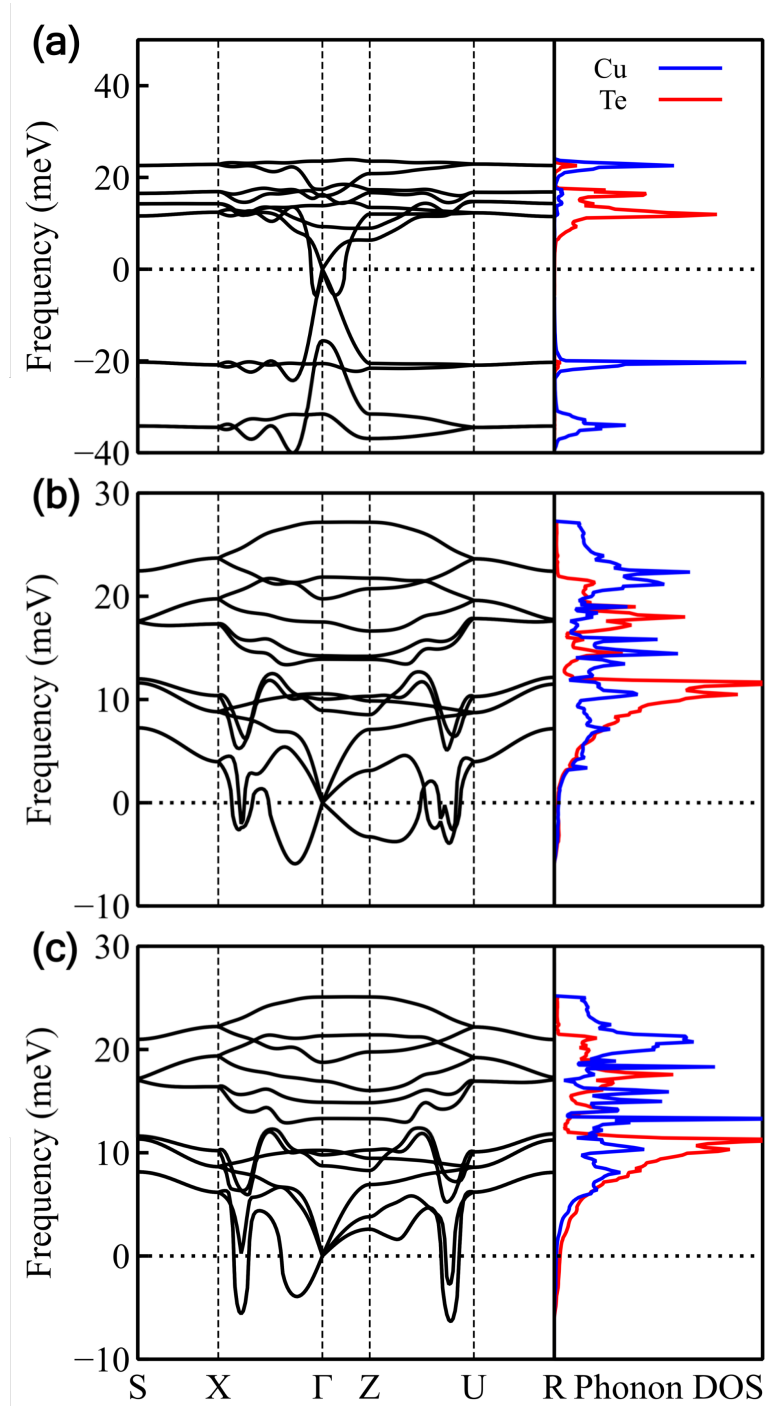

Fig. 3. Phonon dispersion curves and phonon density of states (DOS) of CuTe with (a) SCAN, (b) r²SCAN, and (c) r²SCAN+$U$. The imaginary frequencies signal dynamical instability of the non-CDW structure.

We next explored the lattice dynamics of CuTe using SCAN and r²SCAN. Figure 3 shows phonon dispersion curves and phonon density of states (DOS) of the non-CDW structure obtained with SCAN, r²SCAN, and r²SCAN+$U$. For SCAN, the phonon dispersion exhibits numerous imaginary frequencies as in Fig. 3(a), and SCAN+$U$ also shows similarly unphysical behavior. This indicates that the SCAN functional may not be suitable for phonon calculations in CuTe, where accurate force constants are required. In contrast, the phonon bands obtained with r²SCAN(+$U$) are physically reasonable and show phonon soft modes at $q_{CDW}$=(0.4, 0.0, 0.5). Notably, unlike PBE [12], r²SCAN exhibits the soft modes at $q_{CDW}$ even without $U_{eff}$. This is consistent with the Te-Te bond-length modulation results, where a finite separation between $d_{min}$ and $d_{max}$ is already present at $U_{eff}$=0 eV.

With the inclusion of $U_{eff}$, the phonon soft modes at $q$=(0.4, 0.0, 0.5) and $q$=(0.4, 0.0, 0.0) become more pronounced. The phonon dispersion curve calculated with r²SCAN+$U$ at $U_{eff}$=5 eV is similar to that from the previous PBE+$U$ phonon with $U_{eff}$=9eV, which also reported the imaginary frequencies at $q$=(0.4, 0.0, 0.5) and (0.4, 0.0, 0.0) [12]. The phonon DOS further shows that Cu and Te contributions are dominant at higher and lower frequency ranges, respectively, reflecting the lighter mass of Cu and the heavier mass of Te. Including $U_{eff}$ leads to a decoupling of Cu and Te vibrations like that in the PBE case, with an increased Te contribution and a reduced Cu contribution in the low-frequency region.

In addition, we performed the phonon calculations with SCAN+$U$+rVV10 and r²SCAN(+$U$)+rVV10 after full relaxation. The phonon dispersions with SCAN+$U$+rVV10 still exhibit unphysical features similar to those in Fig. 3(a). The phonon structures with r²SCAN(+$U$)+rVV10 are qualitatively the same as those of r²SCAN(+$U$) in Fig. 3(b,c) with the phonon soft modes at $q$=(0.4, 0.0, 0.5) and $q$=(0.4, 0.0, 0.0).

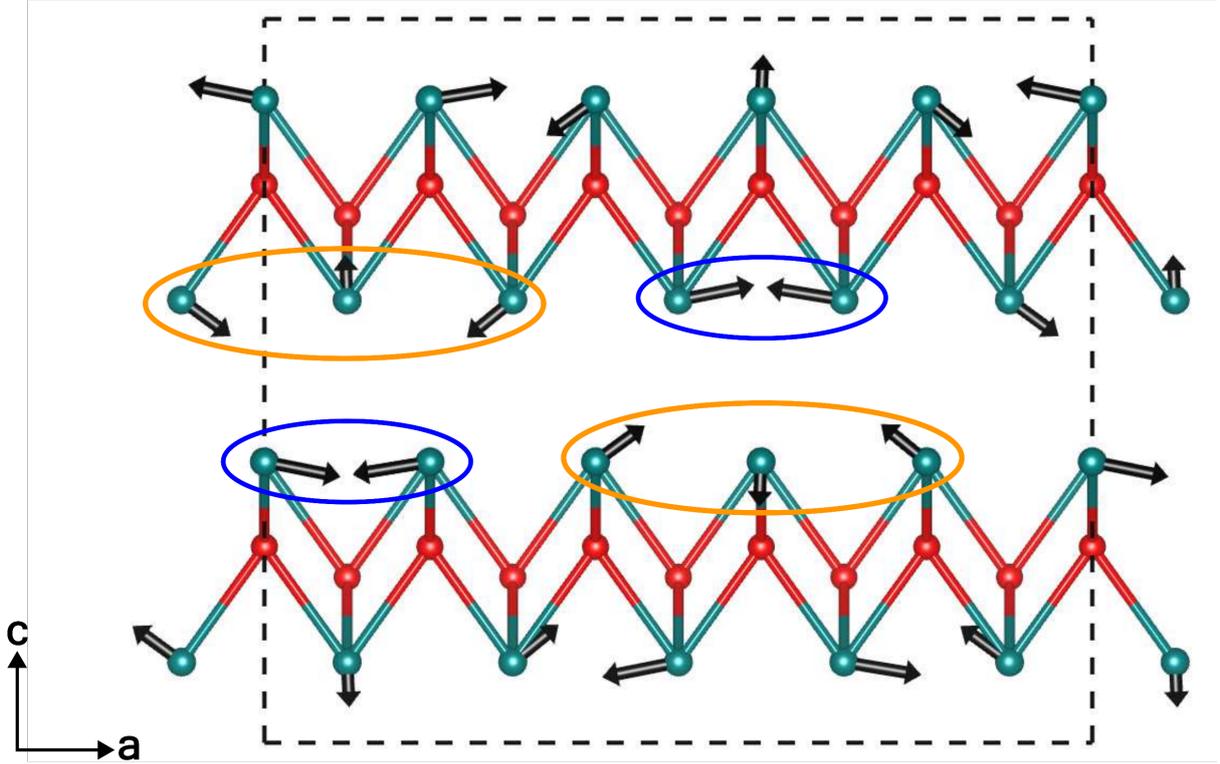

Fig. 4. Lattice displacements of the phonon soft mode at $q=(0.4, 0.0, 0.5)$ with r²SCAN+$U$. Black arrows indicate distortions of Te atoms. Blue and orange circles denote Te dimer and Te trimer, respectively.

Figure 4 illustrates the real-space atomic displacements of the phonon soft mode at $q_{CDW}=(0.4, 0.0, 0.5)$. The eigenvector shows a Te-Te bond-length modulation that produces Te dimers and Te trimers within each layer. Between adjacent layers, dimers and trimers face each other across the van der Waals gap. These features are consistent with the experimental CDW modulation of quasi-1D Te chains [10]. The modulation of Te-Te chains is much stronger along the $a$ direction than along the $c$, while the displacements along the $b$ direction are absent. These results show that r²SCAN+$U$ properly reproduces the experimental CDW instability and the corresponding Te modulation in CuTe.

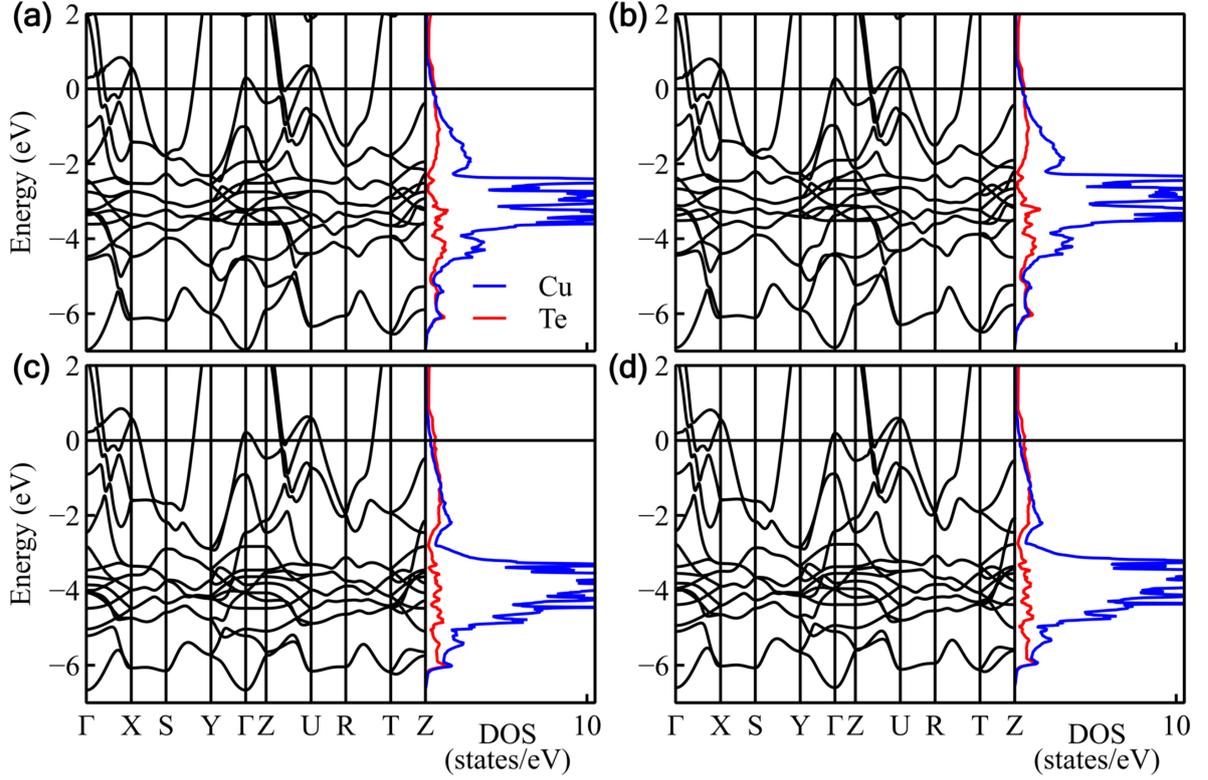

Fig. 5. Band structures and DOS of the non-CDW structure with (a) SCAN, (b) r²SCAN, (c) SCAN+$U$, and (d) r²SCAN+$U$. The Fermi level is represented by horizontal black lines at 0 eV.

We further investigated the electronic band structures and DOS for different functionals and $U_{eff}$ values, as shown in Fig. 5. In contrast to the functional dependence observed in the CDW modulation and phonon dispersions, SCAN and r²SCAN yield almost identical band structures and DOS, both with and without $U_{eff}$. This indicates that the electronic structure is less sensitive to the choice of functional than the CDW instability and lattice dynamics.

With the inclusion of $U_{eff}$, the main peak of Cu DOS shifts to lower energies from ~-3eV to ~-4 eV, and the Te DOS becomes more dominant than Cu DOS near the Fermi level ($E_F$). This suggests the reduced hybridization of Cu and Te at $E_F$, which can enhance the one-dimensional character of the Te-Te chains. In addition, SOC and rVV10 vdW corrections were also tested and found to have an insignificant effect on these electronic structures, consistent with the CDW modulation and phonon results.

Table 1. Calculated lattice constants of the non-CDW phase under various computational settings. $U$ denotes $U_{\text{eff}} = 5$ eV, and $\epsilon$ represents the relative error of relaxed lattice constants with respect to the experimental values [10].

| Functional | $a$ | $b$ | $c$ | $\epsilon_a$ (%) | $\epsilon_b$ (%) | $\epsilon_c$ (%) |
|---|---|---|---|---|---|---|
| PBE | 3.272 | 4.009 | 7.499 | +4.3 | -1.2 | +8.7 |
| SCAN | 3.232 | 3.960 | 7.181 | +3.0 | -2.4 | +4.0 |
| SCAN+rVV10 | 3.193 | 3.949 | 7.073 | +1.8 | -2.7 | +2.8 |
| SCAN+$U$ | 3.135 | 3.988 | 7.183 | -0.1 | -1.8 | +4.1 |
| SCAN+rVV10+$U$ | 3.116 | 3.969 | 6.964 | -0.7 | -2.2 | +0.9 |
| r²SCAN | 3.272 | 3.983 | 7.290 | +4.3 | -1.9 | +5.6 |
| r²SCAN+rVV10 | 3.249 | 3.963 | 7.043 | +3.5 | -2.4 | +0.6 |
| r²SCAN+$U$ | 3.156 | 4.028 | 7.314 | +0.6 | -0.8 | +6.0 |
| r²SCAN+rVV10+$U$ | 3.133 | 4.011 | 7.099 | -0.2 | -1.2 | +2.9 |
| Expt. | 3.138 | 4.059 | 6.902 | - | - | - |

Finally, we fully relaxed the non-CDW structure and analyzed the deviations of the calculated lattice constants from the experimental values [10]. Table 1 summarizes the obtained lattice constants and their relative errors under different computational conditions. Without vdW and $U_{\text{eff}}$ corrections, the errors in the $c$ lattice constant are large (> 4 %), indicating that PBE, SCAN, and r²SCAN all overestimate the interlayer spacing. Among these, PBE yields the largest error, whereas SCAN and r²SCAN are closer to experiment because they already incorporate intermediate range vdW interactions [30,37,41]. For all three functionals, the $a$ lattice constant is overestimated whereas the $b$ lattice constant is underestimated.

With the rVV10 vdW correction, the error in the $c$ lattice constant is significantly reduced. The inclusion of $U_{eff}$ decreases the errors in the $a$ and $b$ lattice constants, with a particularly strong improvement along the $a$ axis, which corresponds to the quasi-1D chain direction. When both vdW and $U_{eff}$ corrections are applied, the errors of all lattice constants become less than 3 %. Under these conditions, both SCAN and r²SCAN provide lattice constants that are comparably accurate relative to experiment. These results further demonstrate that the discrepancy between SCAN and r²SCAN becomes more pronounced in properties that demand high accuracy, such as phonons.

**Conclusions**

In conclusion, we systematically compared SCAN and r²SCAN in describing the CDW instability, lattice dynamics, and electronic structure of CuTe. The Te-Te modulation in the 5×1×2 structure shows that r²SCAN with $U_{eff}$=5 eV reproduces the experimental CDW modulation, whereas SCAN does not reproduce it even at much larger $U_{eff}$. This demonstrates that r²SCAN+$U$ provides a more accurate and robust description of the CDW modulation in CuTe. In the phonon calculations, SCAN produces unphysical dispersions with many phonon soft modes regardless of $U_{eff}$, whereas r²SCAN exhibits physically meaningful phonon bands and correctly captures the phonon soft mode at $q_{CDW}$=(0.4, 0.0, 0.5). The atomic displacements of this soft mode with r²SCAN+$U$ are consistent with the experimentally reported Te dimer-trimer modulation. In contrast, SCAN and r$^2$SCAN yield nearly identical band structures and DOS, and both functionals provide similarly accurate lattice constants. Therefore, our results highlight that r$^2$SCAN offers a significantly more reliable description than SCAN for investigating CDW physics and lattice dynamics in CuTe. These findings provide practical guidance for the use of meta-GGA functionals such as SCAN and r$^2$SCAN in correlated low-dimensional materials.

**Acknowledgements**

We thank Bongjae Kim for helpful discussion. This work was supported by the National Research Foundation of Korea (NRF) (Grant No. 2022R1F1A1063011, RS-2025-00557388) and KISTI Supercomputing Center (Project No. KSC-2025-CRE-0012). This work was also supported by Global-Learning & Academic research institution for Master's·PhD students, and Postdocs (LAMP) Program of the NRF grant funded by the Ministry of Education (No. RS-2023-00301914).

**Author contributions**

S.J. performed the calculations and wrote the first draft. S.K. supervised the project and edited the manuscript. All authors reviewed the manuscript.

**Data availability**

The data supporting the findings of this study are available from the corresponding author upon reasonable request.

**Declarations**

The authors declare no competing interests.